\documentclass[onecolumn]{emulateapj}

\def\etal {{et al.~}}
\def\apj{{\it ApJ}}
\def\mnras{{\it MNRAS}}
\def\araa{{\it ARA\&A}}
\def\nat{{\it Nature}}

\begin{document}

\title{{\bf Binary-Disk interaction II: Gap-Opening criteria for unequal mass binaries}}

\author{Luciano del Valle \& Andr\'es Escala }
\affil{{\small Departamento de Astronom\'ia, Universidad de Chile, Casilla 36-D, Santiago, Chile; ldelvalleb@gmail.com}}
\date{{\small {\it 2013 Acepted October 1; }}}
  \begin{abstract}
We study the interaction between an unequal mass binary with an isothermal circumbinary disk, motivated by the theoretical and observational evidence that after a major merger of gas-rich galaxies, a massive gaseous disk with a SMBH binary will be formed in the nuclear region.  We focus on the gravitational torques that the binary exerts onto the disk and how these torques can drive the formation of a gap in the disk. This exchange of angular momentum between the binary and the disk is mainly driven by the gravitational interaction between the binary and a strong non-axisymmetric density perturbation that is produced in the disk, as response to the presence of the binary. Using SPH numerical simulations we tested two gap-opening criterion, one that assumes that the geometry of the density perturbation is an ellipsoid/thick-spirals and another that assumes a geometry of flat-spirals for the density perturbation. We find that the flat-spirals gap opening criterion successfully predicts which simulations will have a gap on the disk and which simulations will not have a gap on the disk. We also study the limiting cases predicted by the gap-opening criteria. Since the viscosity in our simulations is considerably smaller than the expected value in the nuclear regions of gas-rich merging galaxies, we conclude that in such environments the formation of a circumbinary gap is unlikely.

\vspace{1cm}
    \end{abstract}


\section{Introduction}

A binary embedded in a gaseous disk is a configuration that is repeatedly found in astrophysics at a variety of scales. Some examples are the interaction between planetary rings and satellites (Goldreich \& Tremaine  1982), the formation of planets in protoplanetary disks and their migration (Goldreich \& Tremaine 1980, Ward 1997, Armitage \& Rice 2005, Baruteau \& Masset 2013 ), the evolution of stellar binaries (Shu \etal 1987, McKee \& Ostriker 2007), the interaction of stars and black holes in AGNs (Goodman \& Tan 2004, Miralda-Escude \& Kollmeier 2005, Levin 2007), and the expected interaction of massive black holes (MBHs) binaries at the center of merging galaxies. In all these cases, it is fundamental to have a proper understanding of the main dynamical processes that drives the evolution of a binary-disk system. 

The MBH binaries that interacts with gaseous disks are expected to form in the context of hierarchical structure formation (White \& Frenk 1991, Springel \etal 2005 ). In this scenario, the formation and evolution of galaxies is a complex process, where their final states will be sculpted by a sequence of mergers and accretion events. If the galaxies that are involved in this mergers are rich in gas, there is theoretical (Barnes \& Hernquist 1992, 1996; Mihos \& Hernquist 1996; Barnes 2002; Mayer \etal 2007, 2010) and observational (Sanders \& Mirabel 1996; Downes \& Solomon 1998) evidence that a large amount of the gas on these galaxies will reach the central kilo-parse (Kpc) of the newly formed system. Also, there is observational evidence for the existence of MBH at the center of practically all observed galaxies with a significant bulge (Richstone etal 1998, Magorrian etal 1998, Gultekin etal 2009). Therefore, it is expected that the MBH in the center of each galaxy follow the gas flow and form a MBH binary embedded in a gas environment at the central parsec of the newly formed galaxy, as is shown by a variety of numerical simulations (Kazantzidis \etal 2005, Mayer \etal 2007, Hopkins \& Quataert 2010, Chapon \etal 2011). 

Although numerical simulations suggests the formation of MBH binaries, the only conclusive evidence of pairs of black holes come from the observation of quasar pairs with separations of $\sim 100$ Kpc (Hennawi \etal 2006; Myers \etal 2007, 2008; Foreman \etal 2009; Shen \etal 2011, Liu \etal 2011) and some acreting black holes with separations of the order or smaller than one Kpc (Komosa \etal 2003, Fabbiano \etal 2011, Comerford \etal 2012). On the other hand, there is evidence of at least one MBH binary with a separation of a few parsecs (Rodriguez \etal 2006), but in general, the observational evidence of bound MBH binaries remains elusive, and most of the candidates have observational signatures that can be explained by other configuration and processes different from a MBH binary (Valtonen \etal 2008; Komossa \etal 2008; Boroson \& Lauer 2009; Tsalmantza \etal 2011; Eracleous \etal 2011; Dotti, Sesana \& Decarli 2012 and references therein). 
 
Considering the lack of observational evidence, it is crucial to get more insight of the dynamical process of the binary-disk interaction to determine in what type of merger remnants it is more probable to find these binaries and how the binary separation of these systems depends on the characteristics of the central parsec of the mergers remnants.

A considerable amount of work and progress on the understanding of the interaction of a MBH binary with a gas environment has been made since Escala \etal (2004, 2005) showed, with numerical simulations, that ``when the binary arrives at separations comparable to the gravitational influence radius of the black hole ($R_{\rm inf}=2GM_{\rm BH}/(v^2+c_{\rm s}^2)$)'', localy the binary stars dominate the total gravitational field and the gas tends to follow the gravitational potential of the binary, forming a non-axisymmetric density perturbation that interacts gravitationally with the binary and drives a decrease on the binary separation. Due to the self-similar nature of the gravitational potential, in the regime where the gravitational potential is domianted by the MBH binary, the non-axisymmetric density perturbation is also self-similar in nature. This suggest that, although in the simulations of Escala \etal (2004, 2005) the shrinking of the binary stops at the gravitational resolution, the fast decay will continue down to scales where the gravitational wave emission is effective enough to bring the binary to coalescence.

For systems like the ones explored by Escala \etal (2004, 2005), where the gas mass of the disk is much greater than the binary mass, it is expected that the time the binary takes to merge will be on the order of a few initial orbital times (Escala \etal 2005, Dotti \etal 2006), unless a dramatic change happens in the nearby gas of the binary where the strong non-axisymmetric density perturbation forms (such as gap formation or gas ejection by the BH accretion luminosity). This contrasts with the results of simulations of disks with negligible masses, compared to the binary mass, where the coalescence time is found to be on the order of several thousands of local orbital times (Artymowicz \& Lubow 1994; Ivanov, Papaloizou, \& Polnarev 1999; Armitage \& Natarajan 2002; Milosavljevic \& Phinney 2005), which  for $\rm M_{BH} \geq 10^{7} M_{\odot}$ is even longer than the Hubble time (Cuadra \etal 2009).

This fast/slow  (few orbital times versus several thousand orbital times) migration duality can also be found in simulations of protoplanetary disks that harbour planets (eg. extreme mass ratios; Ward \etal 1989, Ward 1997, Bate \etal 2003, Armitage \& Rice 2005; Baruteau \& Masset 2012, Kocsis \etal 2012). In this type of simulations, the planet/star binary is an extreme mass ratio binary ($q\ll 1$) and the fast/slow migration regime is defined as TypeI/TypeII migration. In the Type I regime, the protoplanetary disk experiences a perturbation, due to the small gravitational potential of the planet. This allows a fast migration of the planet (in the order of a few $t_{\rm orb}$) with a characteristic time scale that scales as the inverse of the planet mass ($t_{\rm migration}\propto M_{\rm p}^{-1}$). The Type II migration (slow migration) is experienced by a planet when its Hill radius is greater than the local pressure scale height of the protoplanetary disk ($R_{\rm Hill}>>h$). In this case the perturbation in the protoplanetary disk due to the presence of the planet becomes important and the planet begins to excavate a gap on the disk. This leads to a coupled evolution of the planet and the disk on a viscous time scale, making the migration time much longer.

For the case of a comparable mass binary embedded in a disk, as in the extreme mass ratio case of a star-planet-disk system, the threshold between the fast and slow migrations is also determined by the formation of a gap or cavity in the disk. Therefore, if we can determine for which systems a gap is opened we can determine in what system a fast or slow migration will occur. For this reason, in a previous work (del Valle \& Escala 2012) we derived a gap-openin criterion that we test with numerical simulations of equal mass binaries embedded in gas disks. In this previous publication, we find that the gap-opening criterion, and hence the threshold between the fast and slow regimes, is determined by the relative strength between the gravitational and viscous torques. In this paper, our aim is to extend the study of a gap-opening criteria to the case of binaries with moderate mass ratio ($0.1\leq q \leq 1$).

This paper is organized as follows. In Section 2 we extend the analytic gap-opening criterion derived in our previous work (del Valle \& Escala 2012) to the case of moderate mass ratio binaries ($0.1 \leq q \leq 1$). In Section 3 we present the setup of the numerical simulations that we use to test the extended analytic gap-opening criterion. In Section 4 we describe how we identified the formation of a gap in our numerical simulations and we test the extended analytic gap-opening criterion against this numerical simulations. In Section 5 we study why the formation of a gap in some simulations is not well predicted by the extended gap-opening criterion and we derived a new gap-opening criterion that is consistent with all our numerical simulations. In Section 6 we study the limits for the final evolution of a binary embedded in a gas disk that are predicted by this successful analytic gap-opening criterion. Finally, in Section 7 we discuss the implication of our results for real astrophysical systems and we present our conclusions.

\section{GAP-OPENING CRITERIA FOR UNEQUAL MASS BINARIES}
The most widely studied case for binary-disk interaction, is the case of a binary embedded in a gas disk with much smaller mass than the mass of the primary. This limiting case of low-mass disk ($M_{\rm disk}/M_{\rm primary}\ll 1$) is typically found in the late stages of star/planet formation (Lin \& Papaloizou 1979; Goldreich\& Tremaine 1982; Takeuchi et al. 1996; Armitage \& Rice 2005; Baruteau \& Masset 2012). In these studies, when the binary has an extreme-mass ratio ($q\ll 1$), the gravitational potential produced by the secondary is treated as a perturbation to the axisymmetric gravitational potential of the primary-disk system, allowing a linear approximation for the equations of motion of the secondary. From this approximation, studies find that the sum of the torques, arising from the inner and outer Lindblad and co-rotation resonances, drive the interaction between the secondary and the disk. This approach leads to predictions of the gap structure that are consistent with simulations within the same regime of validity of it ($q\ll 1$ and $M_{\rm disk}/M_{\rm primary} \ll 1$) (Ivanov et al. 1999; Armitage \& Natarajan 2002; Nelson \& Papaloizou 2003; Haiman et al. 2009; Baruteau \& Masset 2012).

Motivated by the success of this approach in the planetary regime ($q\ll 1$), some authors extrapolate this analysis to other cases where $q\sim 1$ (Artymowicz \& Lubow 1994, 1996; Gunther \& Kley 2002; MacFadyen \& Milosavljevic 2008) regardless of the strong nonlinear perturbation that is produced by the non-axisymmetric gravitational field of the binary, which breaks the validity of the linearization of the equation of motion in this regime ($q\sim 1$) (Shi et al. 2012).

In this paper, we study the case of binaries with moderate mass ratio ($0.1\leq q\leq 1$) interacting with a gaseous disk of comparable mass ($M_{\rm disk}/M_{\rm bin} \sim 1$) without any assumption of linearity. We consider the tidal nature of the binary-disk interaction to model the torques between the binary and the disk instead of a resonant process like the one that appears in the linear approximation. This tidal-torque approach is motivated by the work of Escala \etal (2004, 2005) where they found that the exchange of angular momentum between an equal mass binary and a disk is driven by the gravitational interaction between a strong non-axisymmetric density perturbation on the disk and the equal mass binary. Using this approach, del Valle \& Escala 2012 (hereafter dVE12) successfully test an analytical criterion for the formation of a gap in the case $q=1$, assuming that the gravitational interaction between this strong non-axisymmetric density perturbation and the binary was the main process that drives the exchange of angular momentum between the disk and the binary. We extend our gap-opening criterion to the case of unequal but comparable mass binaries ($0.1 \leq q \leq 1$) for which the non-axisymmetric potential of the binary is still sufficiently strong to drive the formation of a strong global non-axisymmetric density perturbation in the disk.

The shape and size of the strong non-axisymmetric density perturbation is determined by the dominant gravitational potential of the binary, whose typical scale lenght is the binary separation. Therefore, as in dVE12, we can assume that the torque produced by the non-axisymmetric density perturbation onto the binary can be written as $\tau=-a^2\mu\rho G K_q$ where $\rho$ is the density of the perturbation, $a$ is the binary separation which determines the scale lenght of the non-axysimmetric density perturbation, $\mu=m_1m_2/(m_1+m_2)$ is the reduced mass of the binary, and $K_q$ is a parameter that depends on the geometry of the density perturbation. In principle the parameter $K_q$ can depend on the mass ratio of the binary because the shape of the non-axysimmetric density perturbation is determined by the non-axysimmetric gravitational potential of the binary.

To derive the criterion for the opening of a gap in the disk, we follow the same procedure used in dVE12. We compare the gap-opening time scale $\Delta t_{\rm open}$ (determined by the torque that the binary exchange over the disk) with the gap-closing time scales $\Delta t_{\rm close}$ (Goldreich \& Tremaine 1980). The gap formation requires that $\Delta t_{\rm open}<\Delta t_{\rm close}$, therefore it is straightforward to find that the binary will open a gap in the disk if 

\begin{eqnarray}
\frac{\Delta t_{\rm open}}{\Delta t_{\rm close}}&=&\frac{1}{f_{\rm q}}\left(\frac{c_{\rm s}}{v}\right)^3\,\left(\frac{v}{v_{\rm bin}}\right)^2\;\le\; 1 \,,
\label{criterion1}
\end{eqnarray}

where $f_{\rm q}=2K_q/\alpha_{\rm ss}$ with $\alpha_{\rm ss}$ being the dimensionless viscosity parameter of Shakura \& Sunyaev (1973), $v$ the rotational speed of the binary/density-perturbation system, $v_{\rm bin}$ the keplerian velocity of the binary, and $c_{\rm s}$ the sound speed of the gas in the disk. In this form the gap-opening criterion depends on the relative strength between the gravitational and viscous torques ($K_{\rm q}/\alpha_{\rm ss}$), the thermal and rotational support of the disk $(c_{\rm s}/v)$ and the relative strength betweenthe total mass of the system and the mass of the binary $(v/v_{\rm bin})$. 

\section{INITIAL CONDITIONS AND NUMERICAL METHOD}

We rum SPH simulations to study the binary-disk interaction and test the generalization of our gap-opening criterion. All our simulations consist of a coplanar unequal mass binary of mass ratio $q$, initial separation $a_{0}$, and mass $M_{\rm bin}$ embedded on an isothermal and stable ($Q>1$) gas disk of radius $R_{\rm disk}$ and mass $M_{\rm disk}$. In these simulations we use a natural system of units where [mass] = 1, [distance] = 1, and $G = 1$. In these units, we set the initial radius of the disk as $R_{\rm disk}$ = 30 and the mass of the disk as $M_{\rm disk} = 30$ for all the runs. The disk of gas is initialized with the same density profile that we used in dVE12: a surface density that is constant for $R<R_{\rm c}$ and $\propto R^{-1}$ for $R>R_{\rm c}$, where $R_{\rm c}=0.1 R_{\rm disk}=3$ and a vertical density profile that has the functional form $\cosh\left(z/H_{\rm d}\right)$ where $H_{\rm d}$ is constant for $R<R_{\rm c}$ and $\propto R$ for $R>R_{\rm c}$. With this set up the mass of the disk in the inner region ($R<R_{\rm c}$) is $M_{\rm gas}(<R_{\rm c})=1$. 

The parameter space that we explore with our numerical simulations is determined by the variation of four parameters $a_0$, $M_{\rm bin}$, $h_{\rm c}$ and $q$, in the ranges, $a_0\in[2,6]$, $M_{\rm bin}\in[1,33]$, $h_{\rm c}\in[0$.$8,3]$ and $q\in[0.1,1]$. We run 16 simulations (see table 1) with different combinations of these parameters.

Following the numerical setup used in dVE12, we include a fixed Plummer potential (Plummer 1911) with a total mass $\sim 0.12 M_{\rm disk}$. This external potential helps stabilize the disk and will mimic the existence of an external stellar component when we apply our result to the study of SMBH binaries.

{\small
\begin{center}
\centerline{Table 1: Run Parameters}
\vspace{0.4 cm} 
\begin{tabular}{ccccccccccc} \hline 

RUN  && $q$ &&$\rm a_{0}/R_{\rm disk}$ && $\rm M(<r)/M_{\rm bin}$ &&  $\rm (c_{\rm s}/v_{\rm bin})^2$ && $\rm Q_{\rm min}$  \\ \hline \hline 
A1  && 0.1 && 0.100 &&0.0125  &&  3.110 && 2.62 \\
A2  && 0.1 && 0.133 &&0.0222  &&  1.780 && 2.20 \\  
A3  && 0.1 && 0.133 &&0.0133  &&  2.220 && 2.20 \\  
A4  && 0.1 && 0.200 &&0.0500  &&  4.650 && 2.19 \\
  
B1  && 0.3 && 0.100 &&0.0125  &&  1.037 && 2.62 \\  
B2  && 0.3 && 0.133 &&0.0222  &&  0.593 && 2.20 \\  
B3  && 0.3 && 0.133 &&0.0133  &&  0.740 && 2.20\\
B4  && 0.3 && 0.200 &&0.0500  &&  1.550 && 2.19 \\

C1  && 0.5 && 0.100 &&0.0125  &&  0.622 && 2.62\\
C2  && 0.5 && 0.133 &&0.0222  &&  0.365 && 2.20\\
C3  && 0.5 && 0.133 &&0.0133  &&  0.444 && 2.20\\
C4  && 0.5 && 0.200 &&0.0500  &&  0.930 && 2.19\\

D1  && 1.0 && 0.100 &&0.0125  &&  0.311 && 2.62\\
D2  && 1.0 && 0.133 &&0.0222  &&  0.178 && 2.20\\
D3  && 1.0 && 0.133 &&0.0133  &&  0.222 && 2.20\\
D4  && 1.0 && 0.200 &&0.0500  &&  0.465 && 2.19\\
\hline 

\label{TABLA}
\end{tabular}
\end{center}
}

The gaseous disk is modeled with a collection of $5\times10^5$ SPH particles of gravitational softening 0.1. This resolution is $2.5$ times greater that previously used in dVE12. In Appendix B of dVE12, we proved that our conclusions are the same for simulations with $2\times10^5$ SPH particles and $1\times 10^6$ SPH particles. Therefore, the number of SPH particles that we use to model the disk in this paper is large enough to numerically test the gap-opening criterion. For the binary, we use 2 collisionless particles with gravitational softening of 0.1. 

In all simulations the disks are stable. In table 1 we specify the minimum value of the parameter Q of Toomre of each simulation. At the begining of all our simulations, the gravitational potential of the Binary-Disk-Plummer system has a non-negligible non-axisymmetric component, due to the contribution of the binary. Therefore, the disk lacks a well-defined velocity profile $v_{\phi}(r)$. For this reason, we calculate the initial rotational velocity of the system using the same procedure used in dVE12. A symmetric representation of the gravitational potential of the binary is used to compute the initial velocity of the gas. We use the initial orbital radius of the secondary as the radius of this symmetric representation. We refer the reader to the Appendix A of dVE12 for more details of this implementation.

\section{STUDYING THE GAP-OPENING CRITERION FOR UNEQUAL MASS BINARIES}

For testing the analytic gap-opening criterion against the SPH numerical simulations described in the previous section, we first need to define what is a gap. For this purpose we use the same numerical criterion used in dVE12 (see section \S 2 for details) to determine in which of our simulations the binary opens a gap in the disk. We will only outline the key aspects of this numerical criterion. 

To determine if a gap is formed in a certain time $t$ we seek two characteristics: (i) a density peak in the perimeter of the gap whose maximum has to be greater than 0.015 (in internal units of the code) and (ii) that the semi-major axis $a$ of the binary does not decrease by more than $10\%$. With these two characteristics we define a numerical threshold to determine in which simulations a gap is formed. We define as a disk with a gap all disks where condition (i) and (ii) are fulfilled and as a disk without a gap all the disks where neither condition (i) nor (ii) are fulfilled. We will call every simulation where a gap is formed as ${\it opened}$ simulation and every simulation where there is no gap as {\it closed} simulations. The times in which we analyze our simulations are the times in which the binary completes 2, 3, 5, 7, 10 and 15 orbits. For more details on this numerical conditions and their justifications as trait of the gap formation, we refer the reader to dVE12 section \S2.

\begin{figure}[h!] 
\centering
\includegraphics[height=8cm]{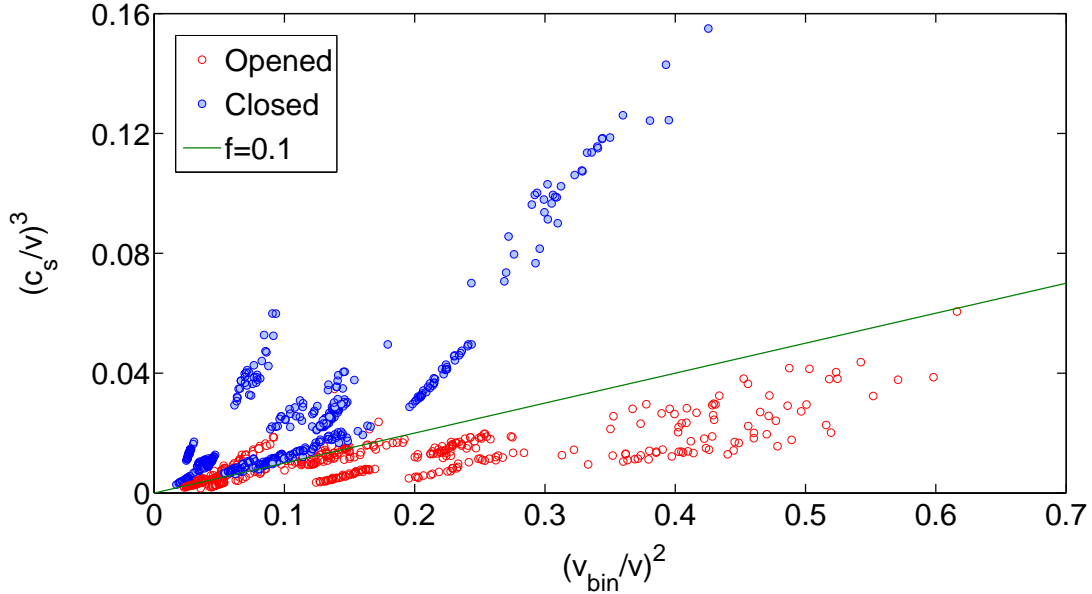}
\caption{Cubic ratio between the sound speed of the gas and the rotational velocity of the binary-disk system $(c_{\rm s}/v)^3$ plotted against the quadratic ratio between the rotational velocity of the isolated \- binary and the rotational velocity of the binary-disk system $(v_{\rm bin}/v)^2$. The red circles are simulations where the binary has opened a gap in the disk ({\it opened} simulations) and the blue filled circles are simulations where the disk does not have a gap ({\it closed} simulations) (see section \S 4). We plot together all the simulations with different values of the mass ratio $q$ that we explore. The green line is the linear $q$-independent threshold between the {\it opened} simulations and the {\it closed} simulations that is predicted by our analytic gap-opening criterion. Below the green line are the {\it opened} simulations and above the line are the {\it closed} simulations. The slope of the interface is the function $f$.}
\label{fig1}
\end{figure}

In figure \ref{fig1}, we plot with open circles the simulations that we called {\it opened} simulations and with filled circles the {\it closed} simulations. In this plot, the horizontal axis is $(v_{\rm bin}/v)^2$ and the vertical axis is $(c_{\rm s}/v)^3$. Each point in the plot corresponds to a given time in which we analyze a simulation. We use the secondary's orbital speed as the speed of the binary/density-perturbation system ($v$) because the strong non-axisymmetric density perturbation is formed by the gas that tends to follow the gravitational potential of the binary and therefore corotates with it.

 We can see, from figure \ref{fig1}, that the group of {\it opened} simulations and the group of {\it closed} simulations populate two different regions of parameter space. The region that is populated by the {\it opened} simulations is the region for which the opening time of a gap is shorter than the closing time ($\Delta t_{\rm open}<\Delta t_{\rm close}$) and the region populated by the {\it closed} simulations is the region for which the closing time of a gap is shorter than the opening time ($\Delta t_{\rm open}>\Delta t_{\rm close}$). Therefore the threshold between these two regions is where the closing time of a gap is equal to the opening time of a gap. We can find the expected shape of this interface evaluating our gap-opening criterion (equation \ref{criterion1}) for the limit case $\Delta t_{\rm open}=\Delta t_{\rm close}$. For this limit, in the parameter space $(\,(v_{\rm bin}/v)^2\, ,\,(c_{\rm s}/v)^3\,)$, the gap-opening criterion predicts that the interface between these two set of simulations has a linear shape of slope $m_{\rm q}=f_{\rm q}=2\,K_{\rm q}\,/\,\alpha_{\rm ss}$.

We first explore if the linear threshold can be assumed to be $q$-independent (i.e. $m_q=f$), even though the geometry of the density perturbation ($K_{\rm q}$) is expected to be $q$-dependent. This assumption implies that in figure \ref{fig1} a line of slope $f$ will be sufficient to model the threshold between the {\it opened} and {\it closed} simulations. We find that the slope of this line is $m=f=0.1$ (green line).  

Figure \ref{fig1} shows that, although the line separates fairly well the distribution of {\it closed} simulations from the distribution of {\it opened} simulations, there are some simulations that are inconsistent with this threshold line. The $17\%$ of the total number of simulations have positions in parameter space that are inconsistent with this $q$-independent threshold line. This discrepancy is even greater for the simulations of mass ratio $q=0.1$ for which $40\%$ of the total number of them has a position in parameter space that can not be explained by this $q$-independent threshold line. This suggest that the value of $f_{\rm q}$ can not be assumed as a $q$-independent one.

\begin{figure}[h!] 
\centering
\includegraphics[height=5.5cm]{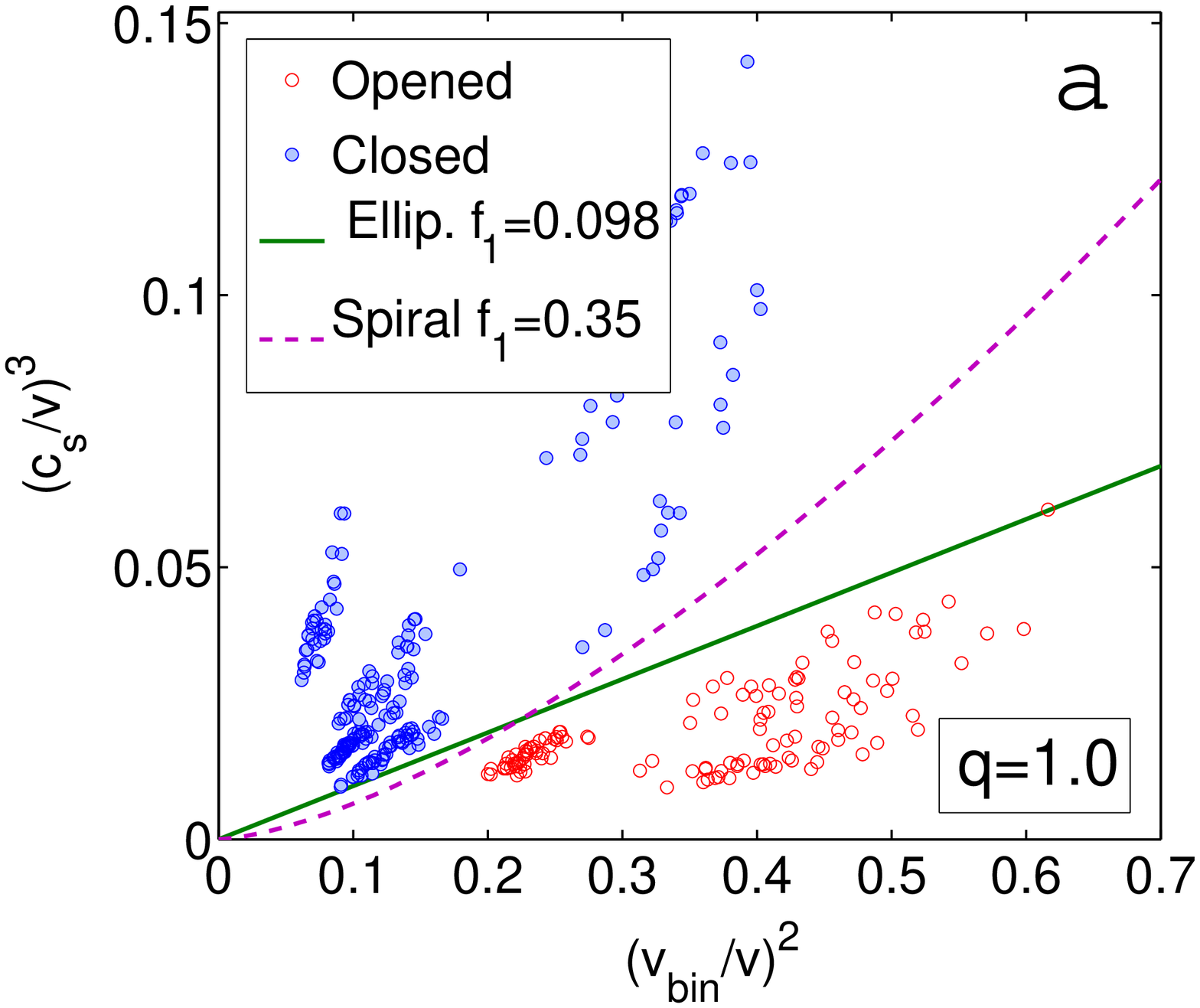}\includegraphics[height=5.5cm]{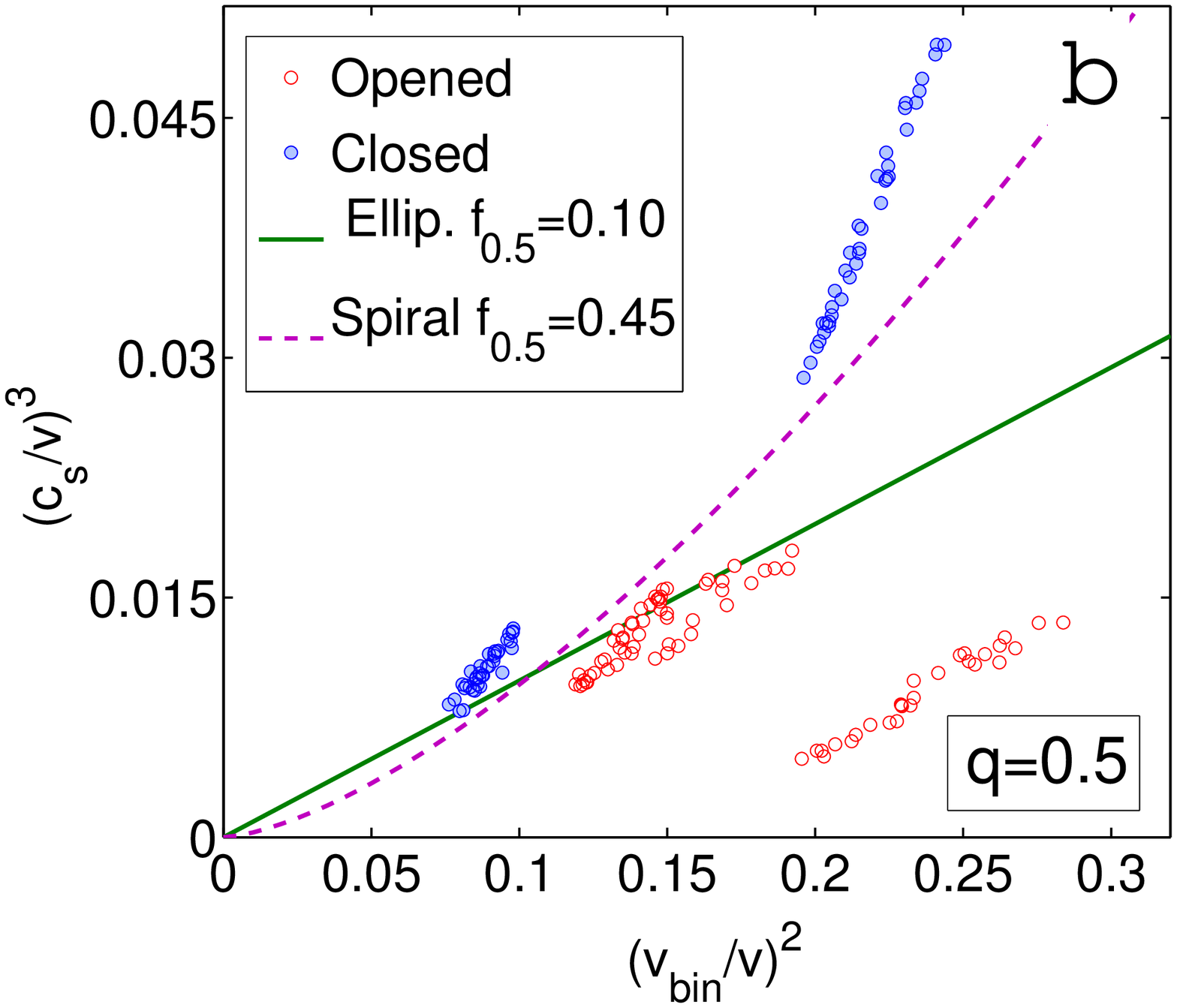}
\includegraphics[height=5.5cm]{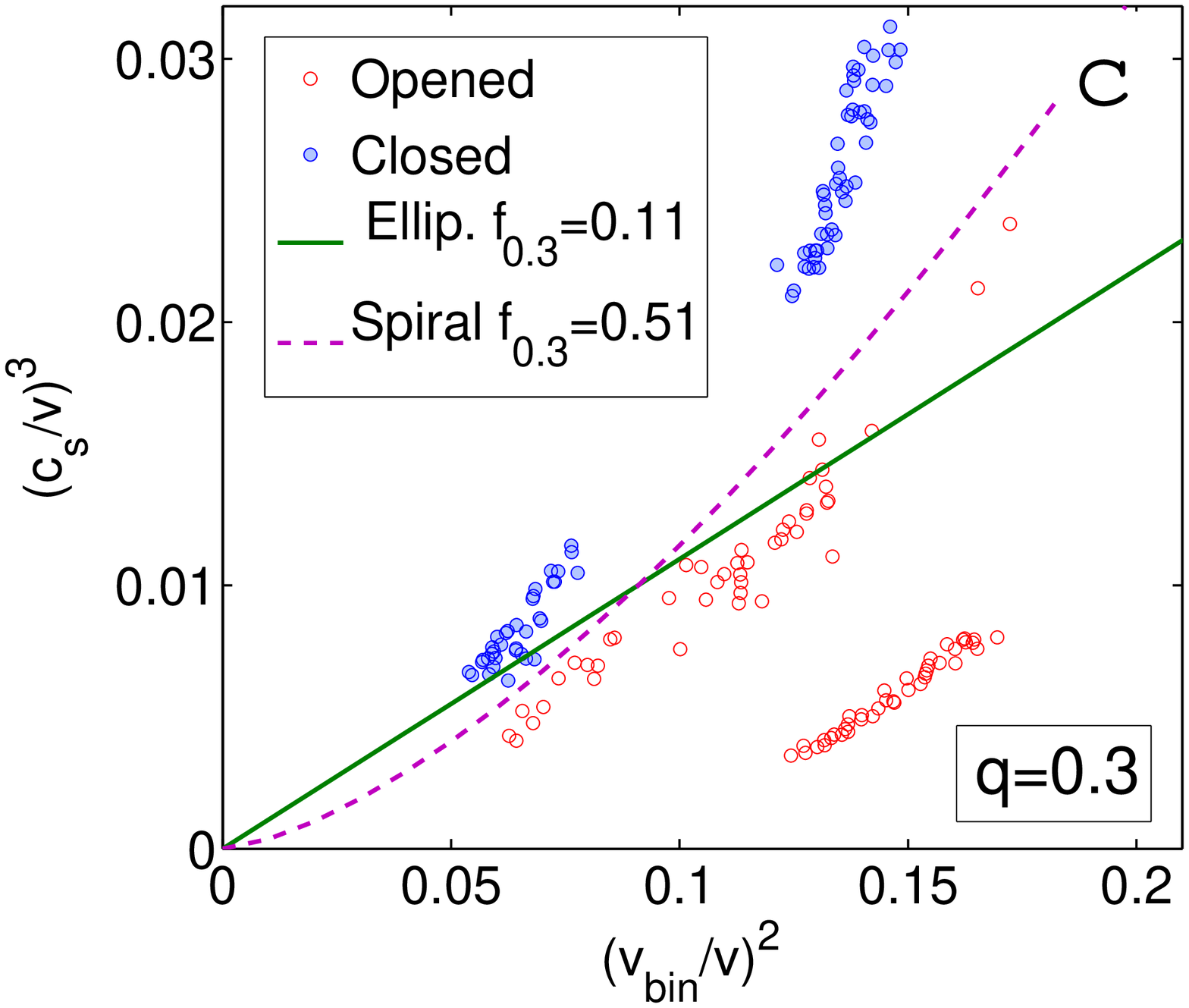}\includegraphics[height=5.5cm]{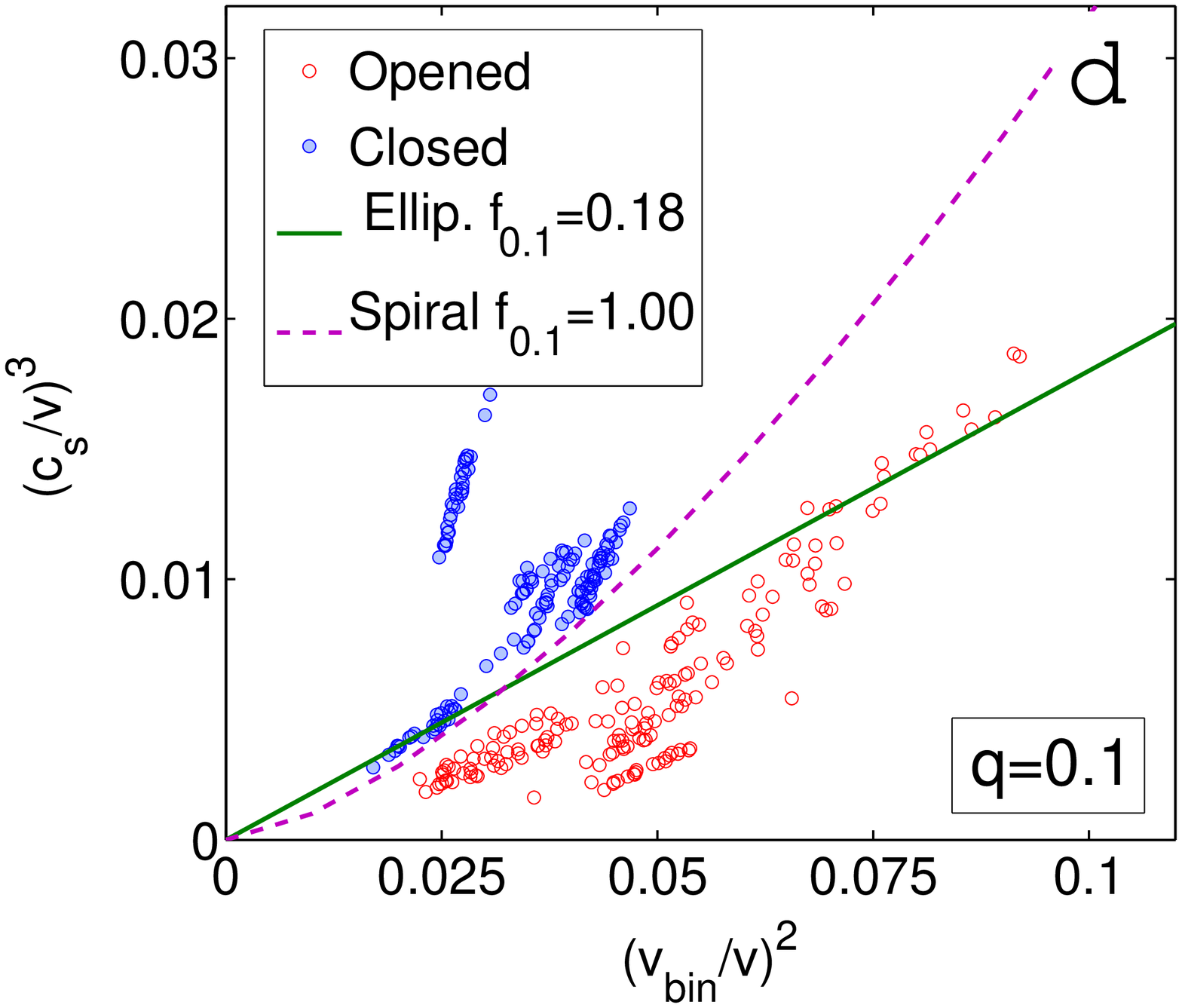}
\caption{Cubic ratio between the sound speed of the gas and the rotational velocity of the binary-disk system $(c_{\rm s}/v)^3$ plotted against the quadratic ratio between the rotational velocity of the isolated binary and the rotational velocity of the binary-disk system $(v_{\rm bin}/v)^2$. The red circles are simulations where the binary has opened a gap in the disk ({\it opened} simulations) and the blue filled circles are simulations where the disk does not have a gap ({\it closed} simulations) (see section \S 4). In all the figures the green line is the threshold between the {\it opened} simulations and the {\it closed} simulations that is predicted by our ellipsoidal gap-opening criterion. The purple dashed curve is the threshold between the {\it opened} simulations and the {\it closed} simulations that is predicted by our flat-spiral gap-opening criterion (see section \S 5). Below the purple dashed curve are the {\it opened} simulations and above the curve are the {\it closed} simulations. From top-left to bottom-right the figures are {\bf 2a}: Simulations with mass ratio $q=1$. {\bf 2b}: Simulations with mass ration $q=0.5$. {\bf 2c}: Simulations with mass ration $q=0.3$. {\bf 2d}: Simulations with mass ration $q=0.1$. For all figures the flat-spiral gap-opening criterion successfully separates the {\it closed} simulations from the {\it opened} simulations (purple dashed curve). In figure {\bf 2a}, for the equal mass binaries ($q=1$), both the flat-spiral gap-opening criterion (purple dashed curve) and the ellipsoidal gap-opening criterion (green line) successfully separate the {\it closed} simulations from {\it opened} simulations.}
\label{fig2}
\label{fig4}
\end{figure}

In figure \ref{fig2} we test if a $q$-dependent slope for the threshold line is a better assumption. We separate the simulations with different values of $q$ in four different plots. In each of this four plots the axis are the same as those in figure \ref{fig1} and we use a different slope $m_q=f_{\rm q}$ for the threshold line. We find that the values of the $q$-dependent slopes that better separates the {\it closed} simulations from the {\it opened} simulation are: $f_{q=1}=0.098$, $f_{q=0.5}=0.100$, $f_{q=0.3}=0.110$ and $f_{q=0.1}=0.180$. We found that the $q$-dependent threshold lines increases the number of simulations that are consistent with the gap-opening criterion, in comparison with the $q$-independent line.

 For the case $q=1$ (figure \ref{fig2}a) we find that all the simulations are well separated by this $q$-dependent linear threshold, in agreement with our previous results (dVE 2012). For the other mass ratios, we find that the $q$-dependent linear threshold separates well almost all our simulations. However, there are still some simulations that are not consistent with this $q$-dependent threshold lines. These deviations are most important for the simulation with mass ratio $q=0.1$ (figure \ref{fig2}d), but even for this extreme case, the number of simulations that deviates from the prediction of our $q$-dependent gap-opening criterion are less than $10\%$. In the next section, we explore the possible causes for these deviations.

\section{DEVIATIONS AND THEIR CAUSES}

In the previous section we test the analytic gap-opening criterion and found that, regardless of it simplicity, it successfully predict the distribution of {\it opened} and {\it closed} simulations in most cases. However, some simulations at certain times, have positions in the space of parameters (figure \ref{fig2}) that are inconsistent with this gap-opening criterion. For example, for $q=0.3$ and $q=0.1$ (\ref{fig2}c and \ref{fig2}d respectively), it is clear that the linear shape of the threshold is not the best curve to explain the separation of {\it closed} and {\it opened} simulations, even if the slope of the linear threshold is not the same for all the values of $q$ that we explore. In order to explain these deviations, we focus on the approximations that we use to derive this gap-opening criterion.

In section \S 2, for the derivation of our analytic gap-opening criterion, we restrict the geometry of the non-axisymmetric density perturbation to an ellipsoid with a scale length equal to the binary separation $a$. For this ellipsoidal geometry the gravitational torque produced on the binary by the density perturbation has a quadratic dependence on the binary separation and can be expressed as $\tau=a^2G\mu K_q$.

 The assumption of an ellipsoidal geometry is based on the work of Escala \etal (2004, 2005). They found that, for the majority of the numerical simulations, the response of the gas to the gravitational potential of a binary has an ellipsoidal shape. However, the numerical simulations of Escala \etal (2004, 2005) were the formation of such ellipsoidal density perturbation is present are far from the regime in which a gap can be formed.

\begin{figure}[h!] 
\centering
\includegraphics[height=7.5cm]{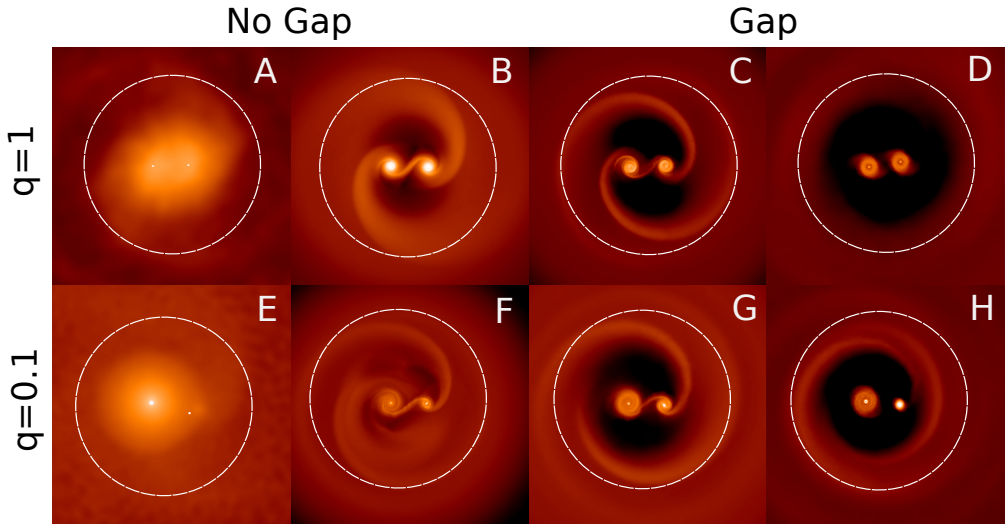}
\caption{Surface density of eight simulations. The white dashed line enclose the region $r<2a$ of the disk, with $a$ the binary separation. The figures from the top
are simulations with $q=1$. The bottom figures are simulatios with $q=0.1$. From left to right the figures show {\bf\ref{fig3}A} and {\bf\ref{fig3}E}: binary that is far from the gap-forming regime, {\bf\ref{fig3}B} and {\bf\ref{fig3}F} binary with parameters in the vicinity of the gap-forming regime that does not form a gap on the disk, {\bf\ref{fig3}C} and {\bf\ref{fig3}G}: binary that begins to excavate a gap on the disk and, {\bf\ref{fig3}D} and {\bf\ref{fig3}H}: binary that excavate a gap on the disk. The geometry of the density perturbation is spiral for the simulations within the gap-forming regime (figures 3C, 3D, 3G and 3H  ) and for the simulation in the vicinity of the gap-forming regime (figures 3B and 3F). In contrast we can see that for the simulation that is far from the gap-forming regime the density perturbation has an ellipsoidaxl geometry for $q=1$ and a pear shape for $q=0.1$ (figures 3A and 3E ).}
\label{fig3}
\end{figure}

In our numerical simulations, we explore the parameter space in the vicinity of the gap-forming regime and we find that the density perturbation has a spiral shape instead of an ellipsoidal one (figure \ref{fig3}). This spiral shaped density perturbations are also found by Escala \etal (2005) in two of their simulations (figure 10 and figure 12), which are in the same gap-forming regime as our simulations.

 The torque produced on the binary by such spiral shaped density perturbation will have the same quadratic dependence on the binary separation $a$, only if the vertical scale of the spiral is comparable to its radial scale (thick-spiral limit). Therefore, the torque given by $\tau\propto a^2$ that we use in the derivation of our analytic gap-opening criterion, will be valid only for the cases where the spiral density perturbation is in the thick-spiral limit.   

The spiral density perturbations that are formed in our simulations tend to be more flat rather than thick, therefore the thick-spiral limit may not be valid for all simulations. For a flat-spiral, the radial scale length of the perturbation is determined by the binary separation $a$ and, as the flat spiral pattern is embedded in the disk, its vertical scale height is truncated by the thickness of the disk ($h_{\rm spiral}\sim H_{\rm disk}$). Therefore the torque produced by this flat-spiral density perturbation can be written as $\tau_{s}=a H_{\rm disk}G\mu K_q$ where the product $a H_{\rm disk}$ is associated to the flat-spiral geometry. 

From the torque produced by a flat-spiral geometry we derive a new gap-opening criterion following the same procedure of section \S 2:

\begin{eqnarray}
\frac{\Delta t_{\rm open}}{\Delta t_{\rm close}}=\frac{1}{f_{\rm q}}\left(\frac{c_{\rm s}}{v}\right)\,\left(\frac{v}{v_{\rm bin}}\right)^2\,\left(\frac{H_{\rm disk}}{a}\right)\;\le\; 1 \,.
\label{criterion2}
\end{eqnarray}

To test if the assumption of a flat-spiral geometry for the density perturbation is a better approximation for our simulations, we compare the shape of the threshold between the group of {\it closed} and {\it opened} simulations that is predicted by the flat-spiral gap-opening criterion (equation \ref{criterion2} in the limit $\Delta t_{\rm open}=\Delta t_{\rm close}$) with the shape of the threshold that is predicted by the ellipsoidal criterion (equation \ref{criterion1} in the limit $\Delta t_{\rm open}=\Delta t_{\rm close}$). 

Assuming for simplicity that $H_{\rm disk}/a=c_{\rm s}/v$ we compare the flat-spiral gap-opening criterion with the ellipsoidal criterion in the parameter space $(\,(v_{\rm bin}/v)^2\, ,\,(c_{\rm s}/v)^3\,)$. In figure \ref{fig4} the purple dashed lines represents the $q$-dependent thresholds between the {\it closed} and {\it opened} simulations, that is predicted by the flat-spiral gap-opening criterion. For this criterion the values of the $q$-dependent parameter $f_{\rm q}$ that better describe the threshold between the {\it closed} and the {\it opened} simulations are: $f_{q=1}=0.35$, $f_{q=0.5}=0.45$, $f_{q=0.3}=0.51$ and $f_{q=0.1}=1.00$.

This threshold has a shape that better separates the {\it closed} from the {\it opened} simulations, compared with the linear threshold predicted by the ellipsoidal gap-opening criterion. In fact, all the simulations that are not consistent with the linear threshold are consistent with the flat-spiral gap-opening criterion. 

For the case of equal-mass binaries (figure \ref{fig4}a) the thresholds predicted by the flat-spiral and ellipsoidal gap opening criteria separated equally good the {\it closed} and {\it opened} simulation. In this case, the parameters of the systems that we explore are in the vicinity of the regime $a\sim H_{\rm disk}$, where the torque associated to the flat-spiral geometry ($\tau \propto a H_{\rm disk}$) and the torque associated to the ellipsoidal geometry ($\tau\propto a^2$) have comparable values making them indistinguishable. On the other hand, in our simulations, the unequal-mass binaries (figures \ref{fig4}b, \ref{fig4}c and \ref{fig4}d) are in systems where the thickness of the disk tends to be smaller than the binary separation, therefore the ellipsoidal torque and the flat-spiral torque have different values. 

\section{LIMITS FOR THE FINAL EVOLUTION OF THE BINARY}

The non-axisymmetric density perturbation that is formed in the disk by the presence of the binary is self-similar in nature and hence when the binary shrinks, the non-axisymmetric density perturbation also shrinks. Therefore, the gravitational interaction between the binary and the non-axisymmetric density perturbation will continue reducing the binary separation unless there is a dramatic change in the nearby gas of the binary such as the formation of a gap.

From the flat-spiral gap-opening criterion we can evaluate how likely it is that this decrease of the binary separation will lead, or not, to the formation of a gap. This is particularly important in the context of the evolution of SMBH binaries where the formation of a gap may stop the shrinking of the SMBH binary at separations where the emission of gravitational waves is not efficient enough to drive the final coalescence of the SMBH.

Assuming for simplicity a disk with a Mestel density profile we can write our flat-spiral gap opening criterion (equation \ref{criterion2}) with its explicit dependence on the binary separation $a$ as

\begin{eqnarray}
  \left(\frac{c_{\rm s}^2H_{\rm disk} }{G M_{\rm bin}}\right)\,\left(\frac{H_{\rm disk}}{q\,a}+\,\frac{H_{\rm disk}}{R_{\rm disk}}\frac{M_{\rm disk}}{M_{\rm bin}}\right)\;\le\; \left(\frac{K_{\rm q}}{\alpha_{\rm ss}}\right)^2 \,,
\label{criterion3}
\end{eqnarray}

where we use the relations: $r_2=(1+q)^{-1}a$, $\mu=qM_{\rm bin}/(1+q)^2$, $v^2_{\rm bin}=G\mu/a$, $v^2=G((M_{\rm gas}(r_2)/r_2)+v^2_{\rm bin}(a/r_2))$ and for a Mestel disk $M_{\rm gas}(r_2)=r_2\,M_{\rm disk}/R_{\rm disk}$. 

From equation \ref{criterion3} we can see that the flat-spiral gap-opening criterion is a decreasing function of $a$ if the disk thickness $H_{\rm disk}$ is constant or does not depend strongly on $a$. In this case the decrease of the binary separation will not drive the system towards the formation of a gap. In fact, the decrease of the binary separation will drive away the binary from the regime where is possible to form a gap.

Although the assumption of a flat-spiral geometry for the density perturbation is more accurate to model the transition from {\it closed} to {\it opened} simulations, the density perturbation on system where $a\ll H_{\rm disk}$ is expected to has an ellipsoidal geometry instead of flat-spiral geometry (for example this can be seen in the simulations of Escala \etal 2004, 2005).  For this reason, we also studied the ellipsoidal gap-opening criterion. Using the same relations that we used to derive the equation \ref{criterion3}, we can write the ellipsoidal gap-opening criterion (equation \ref{criterion1}) with its explicit dependence on $a$ as

\begin{eqnarray}
  \left(\frac{c_{\rm s}^2 H_{\rm disk}}{G M_{\rm bin}}\right)\,\left(\frac{1}{q}\left(\frac{H_{\rm disk}}{a}\right)^3+\frac{H^{3}_{\rm disk}}{a^2\,R_{\rm disk}}\frac{M_{\rm disk}}{M_{\rm bin}}\right)\;\le\; \left(\frac{K_{\rm q}}{\alpha_{\rm ss}}\right)^2 \,,
\label{criterion4}
\end{eqnarray}

We can see from equation \ref{criterion4} that the ellipsoidal gap-opening criterion (like the flat-gap opening criterion) is a decreasing function of $a$ if the disk thickness is constant or does not depends strongly on $a$. Therefore, the decrease of the binary separation will not drive the system towards the formation of a gap. 

In equations \ref{criterion3} and \ref{criterion4} we assume that the disk has a Mestel density profile for which the enclosed mass has the form $M_{\rm gas}(r)=M_{\rm disk}(r/R_{\rm disk})$. If we assume that the disk has a steeper density profile the enclosed mass will be a less steeper function of $r$ or even a decreasing function of $r$. Accordingly, equations \ref{criterion3} and \ref{criterion4} will be more strongly decreasing functions of $a$. Hence, although the assumption of a Mestel density profile for the disk is not based in any expected condition of the density profile in real systems, the results and conclusions derived assuming a Mestel density profile are also valid for any other steeper profile. 

It is important to note that when $a \ll H_{\rm disk}$, we can assume that the disk thickness does not depend on the binary separation because at scales much greater than the binary separation, the binary gravitational potential looks like the gravitational potential of a single object of mass $M_{\rm bin}$. Moreover, the thickness of the disk will be determined by the total mass of the binary+disk and the thermal state of the disk. Therefore, we can conclude that for systems where a gap has never formed, the decrease of the binary separation to the limiting case $a \ll H_{\rm disk}$ will not lead to the formation of a gap.

From this analysis we can conclude that in a large variety of systems a gap will never be opend/formed unless some other process (other than the formation of a gap due to the binary/density-perturbation gravitational interaction) change the distribution of the nearby gas of the binary. This support that is likely that the gas will drive the shrinking of the separation of SMBHs binaries down to scales where the gravitational wave emission is efficient enough to allow their subsequent coalescence. 
\section{DISCUSSION AND CONCLUSIONS}

In simulations of comparable mass binaries ($q\sim 1$) embedded in gas disks, studies find that the time scale for the shrinking of the binary separation can be of the order of a few orbital times (fast migration regime eg; Escala \etal 2004, 2005; Dotti \etal 2006) or of the order of a thousand orbital times (slow migration regime eg; Artymowicz \& Lubow 1994; Ianov \etal 1999; Armitage \& Natarajan 2002; Milosavljvic \& Phinney 2005; Cuadra \etal 2009). The threshold between the fast and slow migration regime for comparable mass binaries, as in the planet migration case (extreme mass ratio binaries $q\ll 1$), is determined by the formation of a gap in the disk. Terefore, in this work, we test a gap-opening criterion that will enable us to estimate in what systems a fast or slow migration will proceed.

In the case of an equal mass binary Escala \etal (2004, 2005) found that, the exchange of angular momentum between the binary and a gaseous disk its driven by the gravitational interaction between the binary and a strong non-axysimmetric density perturbation with an ellipsoidal geometry. Considering this gravitational interaction, in a previous publication (dVE12), we derived a gap-opening criterion that we tested numerically for equal mass binaries. In the present work we study if this ellipsoidal gap-opening criterion is also valid for binaries with moderate mass ratio ($0.1\leq q<1$). For this purpose we ran 12 SPH simulations of binaries with moderate mass ratio embedded in gas disks and we tested the validity of the analytic ellipsoidal gap-opening criterion against this numerical simulations.

We find that the analytic ellipsoidal gap-opening criterion (equation \ref{criterion1}), successfully predicts that the simulations where a gap is formed ({\it opened} simulations) and the simulations where there is no gap on the disk ({\it closed} simulations) are distributed in two separate regions in the parameter space ($(v_{\rm bin}/v)^2$ v/s $(c_{\rm s}/v)^3$) (see figure \ref{fig1}).

 However, there are some simulations, at certain times, with positions in this parameter space that are inconsistent with the ellipsoidal gap-opening criterion (see figure \ref{fig2}). These deviations are most important for the case of binaries with mass ratio $q=0.1$ where roughly the $9\%$ of the total number of these simulations, at certain times, are inconsistent with the ellipsoidal gap-opening criterion.

In our simulations, we find that the strong non-axysimmetric density perturbation has a flat-spiral geometry, instead of the ellipsoidal geometry that we use to derive the ellipsoidal gap-opening criterion (see figure \ref{fig3}). Therefore, we derive a new gap-opening criterion using a flat-spiral geometry for the density perturbation. We find that, this flat-spiral gap-opening criterion (equation \ref{criterion2}) is $q$-dependent and successfully separates the {\it closed} simulations from the {\it opened} simulation. In fact, all the simulations that we explore are consistent with this flat-spiral gap-opening criterion (see figure \ref{fig4}), including the roughly $9\%$ of simulations with mass ratio $q=0.1$ that are inconsistent with the ellipsoidal gap-opening criterion.

The difference between the geometry of the density perturbation that we found in our simulations and the geometry of the density perturbation that are found in the simulations of Escala \etal (2004, 2005) is the result of the different regimes that these two types of simulations explore. In our simulations we explore the vicinity of the gap-forming regime while the simulations of Escala \etal (2004, 2005) are in general far from the gap-forming regime.

Far from the gap-forming regime the gravitational torque that the binary produces on the disk is efficiently absorbed and dissipated through the disk ($\Delta t_{\rm close}\ll \Delta t_{\rm open}$). Therefore, in such regime the gas corotates with the binary in a quasi equilibrium configuration and its structure follow the geometry of the gravitational equipotentials of the binary, which for $q=1$ has an ellipsoidal shape. On the other hand, our simulations have parameters in the vicinity of the gap-forming regime ($\Delta t_{\rm open}\sim\Delta t_{\rm close}$) for which the angular momentum deposited in the gas, through the gravitational torque exerted by the binary, is not efficiently dissipated through the disk and a radial flow of gas is produced. In this non-equilibrium state the density perturbation takes a spiral shape, like the one observed in our simulations within the gap-forming regime (see figure \ref{fig3}) and in others simulations from the literature in the gap-forming regime (Hayasaki \etal 2008, Roedig \etal 2012 ,Shi \etal 2012).

Regardless of the exact geometry of the density perturbation, in the variety of simulations that study the interaction of a comparable mass binary with a gas disk, the torque produced over the binary comes from the same inner region of the disk. For example, for simulations where the density perturbation has an ellipsoidal shape the exchange of angular momentum between the disk and the binary comes from the gravitational interaction between the binary and the ellipsoidal density perturbation which is form in the region of the disk $r\le 2a$ (like is shown in the figure \ref{fig3}A and by the simulations of Escala \etal 2004, 2005). In the simulations where the binary excavates a gap on the disk studies find that the gravitational torque also comes mainly from the inner region $r\le 2a$ and is associated to the gravitational interaction between the binary and transitory streams of gas falling toward the gap region (Roedig \etal 2012, Shi \etal 2012). This can be seen directly from figure 5 of Shi \etal 2012 and figure 9 of Roedig \etal 2012 where they shown the surface torque density on the disk associated to the transitory streams of gas. In our simulations the non-axisymetric density perturbation is also formed in the inner region $r\le 2a$. This can be seen in figure \ref{fig3} were we show for eight simulations the inner region $r\le 2a$ enclosed by a white dashed circle.

From our successfully tested gap-opening criterion, we evaluate if the decrease of the binary separation will lead or not to the formation of a gap. We find that, for a binary embedded in a gas disk with a Mestel density profile (or any steeper density profile) as the binary separation decrease ($a \ll H_{\rm disk}$) the exchange of angular momentum between the binary and the non-axisymetric density perturbation will not lead to the formation of a gap. Then the fast decay of the binary will continue unless some other process changes the distribution of the nearby gas to the binary.

It is important to note that, in the flat-spiral gap-opening criterion (equation \ref{criterion2}) and the ellipsoidal gap-opening criterion (equation \ref{criterion1}), the difficulty that a binary has to open a gap in a disk increase with larger values of the dimensionless viscosity parameter $\alpha_{\rm ss}$ of Shakura \& Sunyaev (1973). For our SPH simulations we estimate that $\alpha_{\rm ss}\approx 0.008-0.016$ from the value of the SPH parameter of artificial viscosity $\alpha_{\rm sph}$ (Artimowicz \& Lubow 1994; Murray 1996; Lodato \& Price 2012; Taylor \& Miller 2012). In massive nuclear disk the gas will be globally unstable and therefore the torques will be larger, with an $\alpha_{\rm ss}$ of order unity (Krumholz \etal 2007, Escala \etal 2007). Moreover, from magnetohydrodynamic (MHD) simulations studies find that the presence of a MHD stresses can significantly increase the torques, with an effective dimensionless viscosity parameter $\alpha\ge 0.2$ (Shi \etal 2012), a factor of $\ge 20$ greater than the estimate value of $\alpha_{\rm ss}$ in our SPH simulations.

From these estimates, we expect that the value of $\alpha_{\rm ss}$ in real gas-rich astrophysical systems, like the nuclear disk in ULRIGs (Downes \& Solomon 1998) and Sub-millimeter Galaxies (Chapman \etal 2003, 2005; Takoni \etal 2006; Swinbank \etal 2010), will be one or two orders of magnitude greater than in our simulations. Therefore, in the nuclear region of the gas-rich merging galaxies, it is more likely that a SMBH binary will not be able to excavate a gap on the gas, allowing the gravitational torques from the gas to shrink the SMBH binary separation down to scales where gravitational wave emission can drive the final coalescence of the binary. 

\vspace{2cm}

{\it Acknowledgements}. L del V research was supported by CONICYT Chile (Grant D-21090518), DAS Universidad de Chile and proyecto anillo de ciencia y tecnología
ACT1101. A.E. acknowledges partial support from the Center of Excellence in Astrophysics and Associated Technologies (PFB 06), FONDECYT Regular Grant 1130458. The simulations were performed using the HPC clusters Markarian (FONDECYT 11090216) and Geryon (PFB 06).



\end{document}